\documentclass[pdfusetitle,aps,reprint,prd,twocolumn,superscriptaddress,nofootinbib]{revtex4-1}
\pdfoutput=1

\usepackage{amssymb}
\usepackage{amsmath}
\usepackage{epsfig}
\usepackage{breakurl}
\usepackage{float}
\usepackage{multirow}
\usepackage{array}
\usepackage{bm}
\usepackage{color}
\usepackage{tikz}
\usepackage[utf8]{inputenc}
\usepackage{braket}
\usepackage{graphicx}
\usepackage{footmisc}
\usepackage{tensor}
\usepackage[normalem]{ulem}
\usepackage{dblfloatfix}
\usepackage[T1]{fontenc}
\usepackage{array}
\usepackage{booktabs}
\usepackage{multirow}
\usepackage{amstext}
\usepackage{multirow}



\usepackage{babel}

\usepackage[force]{feynmp-auto}

\xdefinecolor{mylinkcolor}{rgb}{0,0,0.7}
\usepackage[
	bookmarksnumbered, bookmarksopen, bookmarksopenlevel=2,
	breaklinks=true,
	colorlinks=true, filecolor=mylinkcolor, citecolor=mylinkcolor,
	linkcolor=mylinkcolor, urlcolor=mylinkcolor, menucolor=mylinkcolor,
]{hyperref}
\usepackage{bookmark}

\usepackage{cleveref}
\crefformat{section}{\S#2#1#3} 
\crefformat{subsection}{\S#2#1#3}
\crefformat{subsubsection}{\S#2#1#3}

\usepackage{colortbl}
\definecolor{blue2}{cmyk}{1, 0.1, 0.1, 0}

\definecolor{pyBlue}{RGB}{31, 119, 180}
\definecolor{pyRed}{RGB}{214, 39, 40}
\definecolor{pyGreen}{RGB}{44, 160, 44}
\definecolor{pyBlue2}{RGB}{0, 111, 237}
\definecolor{pyRed2}{RGB}{224, 52, 36}

\definecolor{summersky}{cmyk}{0.71,0.33,0,0.5}
\definecolor{flamingo}{cmyk}{0,0.51,0.71,0.5}
\definecolor{rp}{cmyk}{0.2, 1, 0.6, 0}
\definecolor{pacificblue}{cmyk}{0.95,0.3,0, 0.5}
\definecolor{gray60}{cmyk}{0.4,0.4,0,0.8}

\renewcommand{\[}{\left[}

\renewcommand{\vec}[1]{\mathbf{#1}}

\usetikzlibrary{decorations.markings}
\usepackage{cleveref}
\crefformat{section}{\S#2#1#3}
\crefformat{subsection}{\S#2#1#3}
\crefformat{subsubsection}{\S#2#1#3}

\makeatletter
\def\simgt{\mathrel{\lower$\frac{5}{2}$pt\vbox{\lineskip=0pt\baselineskip=0pt
           \hbox{$>$}\hbox{$\sim$}}}}
\def\simlt{\mathrel{\lower$\frac{5}{2}$pt\vbox{\lineskip=0pt\baselineskip=0pt
           \hbox{$<$}\hbox{$\sim$}}}}

\makeatother

\def\spa#1.#2{\left\langle#1\,#2\right\rangle}
\def\spb#1.#2{\left[#1\,#2\right]}
\def\sand#1.#2.#3{%
\left\langle#1{\vphantom1}\right|{#2}\left|#3\right]}%
\def\sandmp#1.#2.#3{%
\left\langle#1{\vphantom1}\right|{#2}\left|#3\right]}%
\def\sandpm#1.#2.#3{%
\left[#1{\vphantom1}\right|{#2}\left|#3\right\rangle}%
\def\sandmm#1.#2.#3{%
\left\langle#1{\vphantom1}\right|{#2}\left|#3\right\rangle}%
\def\sandpp#1.#2.#3{%
\left[#1{\vphantom1}\right|{#2}\left|#3\right]}%

\renewcommand{\imath}{\mathrm{i}}

\newcommand{\be}{\begin{equation}}
\newcommand{\ee}{\end{equation}}

\def\S{{\mathbb S}}

\allowdisplaybreaks

\begin{document}

\title{Spectral Constraints on Theories of Colored Particles and Gravity}
\author{Aaron Hillman}
\affiliation{Walter Burke Institute for Theoretical Physics, Caltech, Pasadena, CA 91125, USA}
\author{Yu-tin Huang}
\affiliation{Department of Physics and Center for Theoretical Physics, National Taiwan University, Taipei 10617, Taiwan}
\affiliation{Physics Division, National Center for Theoretical Sciences, Taipei 10617, Taiwan}
\author{Laurentiu Rodina}
\affiliation{Beijing Institute of Mathematical Sciences and Applications (BIMSA), Beijing, 101408, China}
\author{Justinas Rumbutis}
\affiliation{Department of Physics and Center for Theoretical Physics, National Taiwan University, Taipei 10617, Taiwan}
\email{\;}

\email{}

\begin{abstract}
In this letter, we consider effective field theories for light fields transforming under the fundamental or adjoint representation of a continuous group. Assuming tree-level completions, we demonstrate that, in the presence of gravity, crossing symmetry combined with two subtraction sum rules leads to constraints on the irreducible representations that the  ultraviolet (UV) degrees of freedom must populate. A spectrum is allowed only if its low energy projection contains the graviton pole. Beautifully, the graviton pole is the anchor of our argument, not an obstruction. Using numerical methods, we also demonstrate that the massless spin-2 must be a singlet under said symmetry group.  
\end{abstract}

\maketitle

\section{Introduction}
Gravitational scattering has long been understood to be highly constrained.  This understanding is an amalgam of everything from sharp results such as the Weinberg-Witten theorem to intuitions and standard lore such as that regarding the fate of global symmetries in quantum gravity. Much of the work on the relation between low energy physics and consistency of quantum gravity comes from the swampland program (see~\cite{Brennan:2017rbf, Palti:2019pca, vanBeest:2021lhn, Grana:2021zvf} for review), largely consisting of conjectures about possible configurations of low energy EFT due to constraints from string theory.  On the other hand, the amplitudes and bootstrap programs have made much progress in recent years constraining low energy EFT from UV consistency on general grounds~\cite{Arkani-Hamed:2020blm,Caron-Huot:2020cmc, Caron-Huot:2021rmr, Guerrieri:2021ivu, Caron-Huot:2022ugt, Caron-Huot:2022jli, Caron-Huot:2024lbf, Tokuda:2020mlf, Alberte:2020jsk, Alberte:2021dnj, deRham:2022gfe, Chowdhury:2021ynh,Bern:2021ppb,Chiang:2022ltp,Beadle:2024hqg, Bertucci:2024qzt, Bellazzini:2023nqj, Bellazzini:2021oaj,Bellazzini:2020cot}, though only very recently approaching the subject of symmetries and consistency of gravitational scattering~\cite{McPeak:2023wmq, Albert:2022oes, Albert:2023jtd, Albert:2023seb}.\\
In this letter, we present the first constraints for scattering in the presence of gravity and gauge or global symmetries on general grounds. This amounts to a constraint on the sets of irreps of the symmetry groups which are permitted to be absent from the UV spectrum, based purely on consistency with Lorentz-invariance, the internal symmetry group, crossing symmetry, and Regge boundedness.  Notably, positivity of the spectral density is not required to derive these equations. In a beautiful irony, the classic $t$-channel graviton pole is a feature of our argument, not a bug.  Ultimately, we constructively demonstrate the following claim: \textit{consistency with gravity at low energies permits only certain patterns of irreps to be absent from the UV spectrum of a theory of colored particles}.\\

We consider the $2\rightarrow 2$ massless scattering amplitude $\mathcal{A}^{i_1 i_2 i_3 i_4}(s,t)$, where $i$'s label the representation of the external state under some symmetry group $\mathcal{G}$ and $s{=}(p_1{+}p_2)^2$, $t{=}(p_1{-}p_4)^2$ are the Mandelstam invariants. On general grounds the amplitude satisfies $\lim_{s\rightarrow \infty}\mathcal{A}^{i_1 i_2 i_3 i_4}(s,t)/|s^2|{=}0 $ for $t{<}0$~\cite{Caron-Huot:2021rmr, Haring:2022cyf}. This implies that one can set up twice subtraction sum rules for the amplitude, which take the form 
\begin{equation}\label{eq: DisDef}
\frac{8\pi G_N}{-t}\delta^{i_1i_4}\delta^{i_2i_3}{+}\cdots=2 \int^\infty_{M^2} \frac{ds'}{s'^3} {\rm}{\rm Im}[\mathcal{A}^{i_1 i_2 i_3 i_4}(s',t)]\,,
\end{equation}
where $\cdots$ are polynomials in $t$ and $M$ is the scale where the massive states enter. If the symmetry is not broken, we can decompose both sides of the above in terms of projectors $\mathbb{P}_{R}$ of a given channel, representing the exchange of states transforming in the $\{R\}$ representation. Note that since the pole from gauge mediator scales as $s^1$, it will not be captured by our twice subtraction sum rules. Therefore in our analysis there is no distinction between global or local symmetry, we only assume that it is not broken. If the symmetry is gauged, we assume that our EFT is well above the confinement scale, far below $M_{pl}$.

The central logic of the argument is simple: we can identify vectors in the space of projectors $\mathbb{P}_R$ which, when contracted, remove all but some chosen set of irreps $\{ \bar R \}$, while keeping the $t$-channel graviton pole on the low energy side.  It is then inconsistent to have only irreps $\{ \bar R \}$ in the UV, as this would require 
\begin{equation}
\frac{8\pi G_N}{-t}{+}\cdots=0\,,
\end{equation}
where $\cdots$ captures polynomial EFT terms. Since the $G_N\neq0$, the above is inconsistent.\footnote{The presence of the graviton pole has also been leveraged to constrain the Regge limit~\cite{deRham:2022gfe,  Noumi:2022wwf}, as well as the other way around~\cite{Haring:2024wyz}. } We demonstrate that this leads to non-trivial constraints in the case of scattering fundamental and adjoint scalars. In particular, in the case of SO(n) adjoint scalars, it is inconsistent to have only singlets and adjoints in the UV. Note that such constraints do not appear for field theory completions, since a massive scalar exchange 
\begin{equation}
\mathcal{A}^{i_1 i_2 i_3 i_4}(s,t)=\frac{\mathbb{P}^{i_1 i_2 i_3 i_4}_R}{s-M^2}{+}{\rm perm}\,,
\end{equation}
is a perfectly UV complete amplitude for any $R$.

These are constraints on the imaginary part of the amplitude in the UV. If we assume perturbative completion, where the amplitude is unitarized while gravity is weakly coupled, then these constraint become a statement on the spectrum of new states. Indeed the scattering of gluons in Heterotic string theory spans all irreps. On the other hand, if the amplitude unitarizes after gravitational loops are incorporated, then as shown in~\cite{Haring:2024wyz}, the eikonal amplitude can reproduce the $t$-channel pole. Expanding the $t$-channel singlet projector onto $s$ and $u$-channel projectors, all irreps are present.

These conclusions are also visible with SDPB numerical bootstrap methods used to bound EFT couplings~\cite{Simmons-Duffin:2015qma, Caron-Huot:2021rmr}. We indeed find that particular irrep configurations must be present in order to have a strictly non-zero $G_N$. Utilizing SDPB also allows us to test scenarios where the massless spin-2 is in some other irrep, and we find that only having it in the singlet gives consistent results.
\section{The dispersive representation for colored S-matrix}
\subsection{Expansion basis for colored amplitudes}
 At low energies, the amplitude is a meromorphic function with simple massless poles, reflecting the exchange of gravitons. We will assume that gravity is unitarized while still weakly coupled, and hence the massless branch cuts from the graviton loops are suppressed. The EFT amplitude can then be organized as
\begin{equation}
    \label{eq: eftamp}
    \mathcal{A}_{\rm Low E}^{i_1 i_2 i_3 i_4} =\mathcal{A}_{\rm Grav}^{i_1 i_2 i_3 i_4}+ B_{\text{poly}}^{i_1 i_2 i_3 i_4}\,,
\end{equation}
where 
\begin{eqnarray}
    \label{eq: Grav}
    \mathcal{A}_{\rm Grav}^{i_1 i_2 i_3 i_4}&=&8\pi G_N\left(\frac{(t-u)^2}{s}\delta^{i_1 i_2}\delta^{i_3 i_4}+{\rm Perm}\right)\,.
\end{eqnarray}
The ``$B$'' functions are polynomials in Mandelstam variables and expanded in suitable color basis, if these operators arises from symmetry preserving UV physics.


For colored amplitudes, the  partial wave expansion is a double expansion in kinematic and color space~\cite{Bachu:2022gof}. In particular, we have 
\begin{equation}
    \label{eq:partialwave}
    \mathcal{A}^{i_1i_2i_3i_4} = \sum\limits_{J, R}n_{J}^{(D)}\, f_{J, R}(s)\mathbb{P}_R^{i_1i_2;i_3i_4}\mathbb{G}_{J}^{(D)}\left(1+\frac{2t}{s} \right)\,,
\end{equation}
where in addition to the expansion in Gegenbauer polynomials $\mathbb{G}_{J}^{(D)}$, we have expanded in the $s$-channel projectors $\mathbb{P}_R^{i_1i_2; i_3 i_4}$ which are the set of orthogonal, irreducible tensors in the indices $i_i$, corresponding to the exchange of irreducible representations $R$. More details can be found in~\cite{Cvitanovic:2008zz}. For SO(n) and SU(n) the number of independent projectors, denoted as $q$, are $n$-independent. For SO(n), $q{=}3$ for fundamental $i_i$'s, corresponding to the singlet, symmetric-traceless and the anti-symmetric. In the $s$-channel they take the form:
\begin{align}
  \mathbb{P}_1^{i_1 i_2; i_3 i_4} &= \frac{\delta^{i_1 i_2} \delta^{i_3 i_4}}{n}, \nonumber\\
  \mathbb{P}_{2}^{i_1 i_2; i_3 i_4} &= \frac{1}{2} \left( \delta^{i_1 i_4} \delta^{i_2 i_3} + \delta^{i_1 i_3} \delta^{i_2 i_4} 
    - \frac{2}{n} \delta^{i_1 i_2} \delta^{i_3 i_4} \right), \nonumber\\
  \mathbb{P}_{3}^{i_1 i_2; i_3 i_4} &= \frac{1}{2} \left( \delta^{i_1 i_4} \delta^{i_2 i_3} 
    - \delta^{i_1 i_3} \delta^{i_2 i_4} \right)\,.
\end{align}
For SU(n), we consider $i_1\bar{i_2}i_3\bar{i_4} $ scattering in which we have two irreps, adjoint and singlet, in the $s$ (or $t$) channel:
\begin{align}
  \mathbb{P}^{i_1 i_3}_{1\bar{i_2}\bar{i_4}} &= \delta^{i_1}_{\bar{i_4}} \delta^{i_3}_{\bar{i_2}}-\frac{\delta^{i_1}_{\bar{i_2}} \delta^{i_3}_{\bar{i_4}}}{n}, \quad
  \mathbb{P}^{i_1 i_3}_{2\bar{i_2}\bar{i_4}} &= \frac{\delta^{i_1}_{\bar{i_2}} \delta^{i_3}_{\bar{i_4}}}{n},
\end{align}
and two different irreps, antisymmetric and symmetric, in $u$ channel:
\begin{align}
  \mathbb{P}^{i_1 i_3}_{3\bar{i_2}\bar{i_4}} &= \frac{\delta^{i_1}_{\bar{i_4}} \delta^{i_3}_{\bar{i_2}}-\delta^{i_1}_{\bar{i_2}} \delta^{i_3}_{\bar{i_4}}}{2}, \quad
  \mathbb{P}^{i_1 i_3}_{4\bar{i_2}\bar{i_4}} &= \frac{\delta^{i_1}_{\bar{i_4}} \delta^{i_3}_{\bar{i_2}}+\delta^{i_1}_{\bar{i_2}} \delta^{i_3}_{\bar{i_4}}}{2}.
\end{align}

For adjoint external states $q{=}6$ and $7$ for SO(n) and SU(n) respectively. We present the explicit form for SO(n) in supplementary material ~\ref{sup: Color}. We simply note that the first and fifth operator correspond to the singlet and adjoint representation respectively.

\subsection{Twice subtraction sum rules}
We now derive the dispersive representation for the low energy EFT coefficients. 
Starting with the standard contour in the complex $s$ plane and assuming that the integral at infinity with two subtractions vanishes:
\be \label{starting}
 \oint_{\infty} \frac{ds'}{2\pi i(s'-s)} \frac{\mathcal{A}^{i_1i_2i_3i_4}(s',t)}{s'(s' + t)} = 0 \, ,
\ee
using $s$, $u$ crossing symmetry $\mathcal{A}^{i_1i_2i_3i_4}({-}s'{-}t,t)=\mathcal{A}^{i_1i_3i_2i_4}(s',t)$, \eqref{starting} can be written as
\be \label{eq:disp3}
\begin{split}
 &\left(\text{Res}_{s'=0} +\text{Res}_{s'=-t}+\text{Res}_{s'=s}\right)  \frac{\mathcal{A}^{i_1i_2i_3i_4}(s',t)}{(s'-s)s'(s' + t)}=\\&\int_{M^2}^{\infty} \frac{ds'}{\pi s'(s' + t)} \left(\frac{\text{Im} \mathcal{A}^{i_1i_2i_3i_4}(s',t)}{(s'-s)}+ \frac{\text{Im} \mathcal{A}^{i_1i_3i_2i_4}(s',t)}{(s'+t+s)}\right).
\end{split}
\ee
 Now we substitute \eqref{eq: eftamp} and \eqref{eq:partialwave} in the equation above:
\be \label{eq:disp5}
\begin{split}
&\frac{ B_{\text{poly}}^{i_1 i_2 i_3 i_4}(s,t)}{s(s+t)}-\frac{ B_{\text{poly}}^{i_1 i_2 i_3 i_4}(0,t)}{st}\\&+\frac{ B_{\text{poly}}^{i_1 i_2 i_3 i_4}({-}t,t)}{(s+t)t}-\frac{8\pi G n}{t} \mathbb{P}_1^{i_2i_3;i_4i_1}=\\&n_{J}^{(D)}\, \sum_{JR}\int_{M^2}^{\infty} \frac{ds'}{\pi s'(s' + t)} \text{Im} [f_{J, R}(s')]\\& \times\left(\frac{\mathbb{P}_R^{i_1i_2;i_3i_4}}{(s'-s)}+ \frac{\mathbb{P}_R^{i_1i_3;i_2i_4}}{(s'{+}t{+}s)}\right)\mathbb{G}_{J}^{(D)}(1{+}2t/s')\,.
\end{split}
\ee
Despite appearance, the LHS of the above is analytic except for the $t$-channel graviton pole in the singlet channel $\mathbb{P}_1^{i_2i_3;i_4i_1}$. Note that if the symmetry is gauged, there would be a massless vector pole that scales as $s^1$. This will not be captured by the twice subtraction sum rules, and thus here there is no distinction between global and gauge symmetry in the following discussion.

\subsection{Color Projectors}

To solve eq.(\ref{eq:disp5}), the color factors should be cast into a common basis. We can simply use the same basis we use for the IR amplitude. Denoting $\mathbb{P}_R^{i_1i_2;i_3i_4}=\mathbb{P}_R^{s}$, $\mathbb{P}_R^{i_1i_3;i_2i_4}=\mathbb{P}_R^{u}$, we introduce the conversion matrix,
\begin{equation}
\vec{\mathbb{P}}^{s}=M_{s,{\rm IR}} \vec{\mathbb{P}}^{\rm IR},\quad \vec{\mathbb{P}}^{u}=M_{u,{\rm IR}} \vec{\mathbb{P}}^{\rm IR}\,,
\end{equation} 
where $\vec{\mathbb{P}}^{\rm IR}$ is the color basis for the IR amplitude. 
For example, for SO(n) fundamental representation, we can choose $\vec{\mathbb{P}}^{\rm IR}=\vec{\mathbb{P}}^{t}$ and $M_{st}$ is given by:
\be
\begin{pmatrix}
\frac{1}{2} {-}\frac{1}{n} & \frac{(n{-} 1) (2 {+} n)}{2 n} & -\frac{2 {+} n}{2 n} \\
\frac{1}{n} & \frac{1}{n} & \frac{1}{n} \\
-\frac{1}{2} & \frac{1}{2} ( n{-}1) & \frac{1}{2}
\end{pmatrix}\,,
\ee
where the first, second and third columns correspond to the singlet, symmetric and anti-symmetric representation respectively. $M_{ut}$ is the same as $M_{st}$ with an additional minus sign in the third column. For SO(n) Adjoint representation, the $6\times 6$ matrices are given in \eqref{eq:Mst} and \eqref{eq:Mut}. For SU(n) the projectors no longer span the full space, so we use $(\delta^{i_1}_{\bar{i_2}} \delta^{i_3}_{\bar{i_4}},\delta^{i_1}_{\bar{i_4}} \delta^{i_3}_{\bar{i_2}})$  as the basis of the color factor space. In this basis the $s$ and $u$ projectors are given respectively as
\be
\begin{pmatrix}
-\frac{1}{n} & 1 \\
\frac{1}{n} & 0 \\
0 & 0 \\
0 & 0
\end{pmatrix},\quad \begin{pmatrix}
0 & 0 \\
0 & 0 \\
-\frac{1}{2} & \frac{1}{2} \\
\frac{1}{2} & \frac{1}{2}
\end{pmatrix}.
\ee

\section{Implications of gravity}
Crucially, after projecting on to independent basis eq.(\ref{eq:disp5}) leads to $q$ equalities, one for each basis element. The RHS of eq.(\ref{eq:disp5}) now takes the form 
\begin{equation}
 \sum_{J}\int_{M^2}^{\infty}ds' \text{Im} [\vec{f}_{J}(s')]\left(M_{st}h_{s,J}(s',t)+ M_{ut}h_{u,J}(s',t)\right)\,,
\end{equation}
where each $\vec{f}_{J}(s')$ is a $q$-dimension vector. Now consider a $q$-dimension vector $v=(1,\cdots)$ such that
\begin{equation}\label{eq: NullSpace}
\boxed{\quad\left[M_{st}v\right]_i=\left[M_{ut}v\right]_i=0\,, \quad}
\end{equation}
where $i$ is one of the $q$ irreps. Basically, one is considering the complement to the $i$-th projector. If such a vector exists, then when multiplied to both sides of eq.(\ref{eq:disp5}), after converted into the $\vec{\mathbb{P}}^t$ basis, we arrive at 
\begin{align}
\frac{8\pi G_N}{-t}{+}Poly(t)&=\sum_{j\neq i}\sum_{J}\int_{M^2}^{\infty}ds' \text{Im} [f_{J,j}(s')]\nonumber\\
&\left([M_{st}v]_{j}h_{s,J}(s',t)+ [M_{ut}v]_jh_{u,J}(s',t)\right)\,.
\end{align}
The fact that the first component of $v$ is 1 guarantees that the graviton pole will appear on the LHS. Importantly, if we set $f_{J,j}=0$ for all $j\neq i$, the RHS vanishes and we reach an inconsistency since the gravitational coupling is non-zero! This immediately rules out such spectrum for gravity.

\subsection{Fundamental and Adjoint matter}
Let us start with the case of \textbf{fundamental} representation. We find that for each irrep we can find its corresponding vector $v$, which is listed in table \ref{tab:combined}. Importantly, the vector $v$ exists for all $n$! Therefore we see that fundamental matter in SO(n) or SU(n) demands a UV completion of gravity with at least two irreps! \footnote{It might seem that $SO(2)$ or $U(1)$ is a contradiction, since in \cite{McPeak:2023wmq} it was shown that it is allowed to have a UV completion of fundamental representation scattering ($U(1)$ charge $\pm$1) with just charge 0 exchanges. However in this case there are still two distinct charge 0 exchanged states (parity even and parity odd) which in $SO(2)$ language correspond to two distinct representations.  }

\begin{table}[h!]
\centering
\begin{tabular}{|c|c|c|}
\hline
Group & Irreps & $\vec{v}$ \\
\hline
\multirow{3}{*}{SO(n)} & $\{1\}$ & $\left( 1,\frac{2-n}{(n-1) (2 + n)}, 0 \right)$ \\
\cline{2-3}
& $\{2\}$ & $\left( 1, -1, 0 \right)$ \\
\cline{2-3}
& $\{3\}$ & $\left(1, \frac{1}{n-1}, 0 \right)$ \\
\hline
\multirow{4}{*}{SU(n)} & $\{1\}$ & $\left( 1, \frac{1}{n} \right)$ \\
\cline{2-3}
& $\{2\}$ & $\left( 1,0 \right)$ \\
\cline{2-3}
& $\{3\}$ & $\left( 1, 1 \right)$ \\
\cline{2-3}
& $\{4\}$ & $\left(  1, -1 \right)$ \\
\hline
\end{tabular}
\caption{The vectors $v$ that satisfies eq.(\ref{eq: NullSpace}) for each irrep in the exchange of SO(n) and SU(n) fundamentals.}
\label{tab:combined}
\end{table}

For \textbf{adjoint} matter, we can similarly show that one needs at least three irreps for general $n$. In fact the only allowed UV completion with three irreps is the singlet, adjoint and antisymmetric representation (1,5,6) in our notation~\ref{eq: projbasis}. The vectors $v$ that eliminate all other three irrep completion are given in \ref{tab:3 reps} . To complete the claim that we cannot have completions with only two representation, we just need to show that two representation subsets of ${1,5,6}$ are forbidden. The corresponding vectors are listed in Table \ref{tab:2 reps}. Some four representations can also be removed. For $n=4$ we are able to find $v$ for
$\{1, 2, 4, 6\}$ and $\{1, 3, 4, 5\}$, shown in Table \ref{tab:4 reps}. 

\begin{table}[h!]
\centering
\begin{tabular}{|c|c|}
\hline
Irreps & $\vec{v}$ \\
\hline
$\{1, 5\}$ & $\left( 1, 0, \frac{-4 + n}{3}, \frac{1 - n}{3}, 0, 0 \right)$ \\
\hline
$\{1, 6\}$ & $\left( 1, \frac{4 - 2 n - 3 n^2 + n^3}{2 (-2 + n)}, -\frac{(-3 + n) n^2}{2 (-2 + n)}, 0, 0, 0 \right)$ \\
\hline
$\{5, 6\}$ & $\left( 1, -1 + n, \frac{-3 + n}{2} n, 0, 0, 0 \right)$ \\
\hline
\end{tabular}
\caption{The vectors $v$ that satisfies eq.(\ref{eq: NullSpace}) for two irreps of SO(n).}
\label{tab:2 reps}
\end{table}

Importantly, the spectrum with only singlets and adjoints is ruled out. That is, the UV completion must include irreps outside of that of the low energy states. Let us compare this to a scenario where we can access the massless pole from the gauge boson. This would require one-subtraction which can be justified for specific setups~\cite{Haring:2022cyf, Albert:2023seb}.  Then we can repeat the same analysis, and find that all single irrep completions are ruled out \textit{except} for the adjoint, see Table \ref{tab:1subtr adj}. Thus without the graviton pole, a gauge theory can be UV completed in a self-contained manner.

\section{Graviton poles as other irreps }
\begin{figure}[h] 
   \centering
   \includegraphics[width=3in]{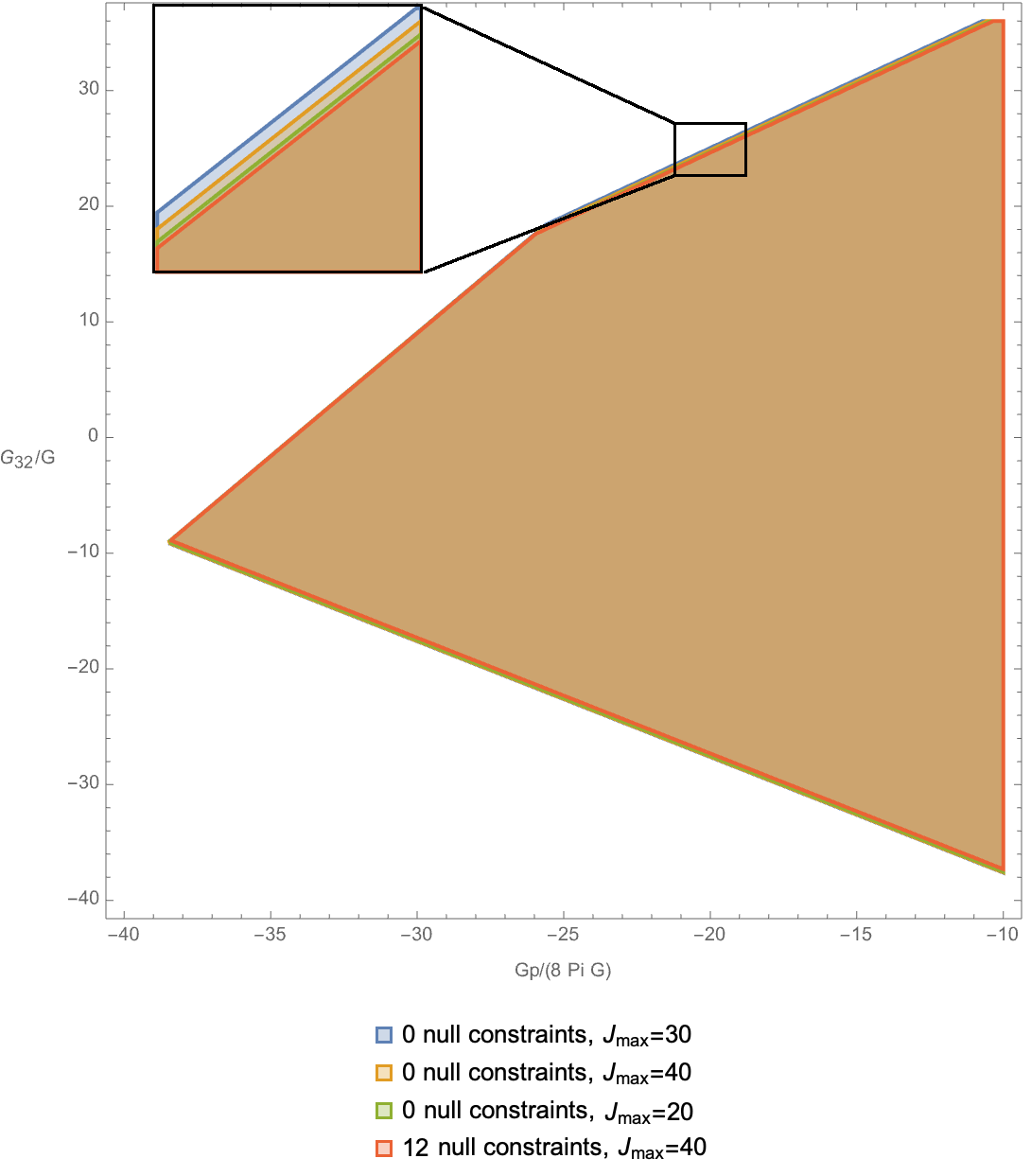} 
   \caption{$G_{32}/(8 \pi G)$ vs $G_p/(8 \pi G)$ 
 with graviton in singlet representation in $d=5$. The bounds seem to be already convergent with low spin truncation. The null constraints here refer to forward limit ($t=0$) null constraints. We find that if the graviton is in some other representation than singlet, there is no allowed region.}
   \label{fig:bound}
\end{figure}

One might be tempted to ask what would happen if the spin-2 massless pole was in some other irrep. While from the four-particle test one can show that a self-interacting massless spin 2 state must be the unique graviton~\cite{Benincasa:2007xk,McGady:2013sga}, here we can show the same even without assuming the spin 2 self-interacts. For this we use a numerical approach, which requires the improved dispersion relation containing only finite number of EFT couplings~\cite{Caron-Huot:2021rmr}. Schematically it is given as
\begin{equation}
\label{eq:disp improved sch0}
\boldsymbol{\mathcal{C}}^{\rm improved}_{\text{low E},t} = \sum_R\left<\boldsymbol{\mathcal{C}}^{\rm improved}_{\text{high E},t}[m^2,J,R]\right>_R \,,
\end{equation}
where the left hand side contains a finite number of couplings multiplied by powers of $t$ packaged in $c^{i_1i_2i_3i_4}(t)$
\be 
c^{i_1i_2i_3i_4}(t)+8\pi G  \frac{\mathbb{P}_R^{i_2i_3;i_4i_1}}{t}\,,
\ee
and the right hand side is a heavy average over the massive states defined in (\ref{sum}). The derivation of the improved dispersion relation is in Appendix \ref{sec:impr disp}. We supplement the improved dispersion relation with null constraints derived for higher powers of $s$ terms of \eqref{eq:disp5}. We sample the equations over spins, masses (including impact parameter region, see \cite{Caron-Huot:2021rmr}) and representations, require the positivity of $\text{Im} [f_{J, R}(s')]$ in \eqref{eq:disp5}, and we search for numerical bounds using the SDPB package \cite{Simmons-Duffin:2015qma}. Here we assume that the spin-2 field can be in any representation $R$ with the $t$ projector given by $\mathbb{P}_R^{i_2i_3;i_4i_1}$. We optimize the following combination of the couplings: 
\be
\begin{split}
G_p &= \frac{1}{4} \left( 3 g_{2, 0} + 7 g_{2, 2} + 4 G_{2, 0} + 26 G_{2, 2} \right).
\end{split}
\ee  
chosen such that it is positive when $G=0$:
\be
\begin{split}
G_p =& \bigg< \frac{1}{m^2}\bigg>_1+\bigg< \frac{5 n-16 }{8 n m^2}\bigg>_2+ \bigg<\frac{25 - 23 n + 6 n^2}{4 (n-2) (n-1) m^2}\bigg>_3 \\ &+\bigg<\frac{3}{8 m^2}\bigg>_4+ \bigg<\frac{1}{8 m^2}\bigg>_5+ \bigg<\frac{7 n-16}{8 (n-2)m^2}\bigg>_6 >0\,,
\end{split}
\ee  
divided by $G$, i.e. we obtain the minimum and maximum value of $G_p/G$. We also plot a 2d region, with $G_{3,2}$ on the vertical axis, in figure \ref{fig:bound}. We find that a solution space only exists for $R=1$, the singlet representation. A more comprehensive study of such bounds will be presented elsewhere.

\section{Conclusions}
\label{sec:conclusions}
We have shown that the presence of the graviton pole, combined with its coupling to colored states, imposes sharp constraints on the UV spectrum of Regge bounded, crossing symmetric amplitudes. This constraint comes in the form that new irreps must be included in the spectrum. It will be interesting to show that such new states cannot come in parametrically large above the cut-off. It is interesting to consider the Completeness Hypothesis  conjecture in our setup, which states that all irreps must be present in the spectrum~\cite{Polchinski:2003bq,Banks:2010zn,Harlow:2018tng,Heidenreich:2021xpr,Rudelius:2020orz}. Since we see that starting with the fundamental or adjoint, consistency requires the presence of a new irrep. The scattering of the later will presumably require further new irreps. A key question is whether the only possible outcome of this iterative process is the presence of all irreps. On the other-hand, we see for scalar $SO(n)$ adjoint scattering that the minimal set of UV irreps consists of a singlet, adjoint and antisymmetric representation. It will be interesting to see if such spectrum can be realized with an explicit UV model.

The presence of higher spin states, with $J>1$, will lead to new constraints from crossing symmetry and positivity. These extra constraints are needed for obtaining two-sided bounds on EFT couplings~\cite{Caron-Huot:2020cmc, Bellazzini:2020cot, Tolley:2020gtv}, and for scattering of identical scalars imply the presence of an infinite number of higher spins~\cite{Chiang:2021ziz}, when any $J\geq2$ are present. This latter result follows simply from analyzing the roots of spin polynomials in the null constraints. It will be interesting to study whether such arguments constrain the allowed irreps that can appear, including when gravity is absent. 


\section*{Acknowledgments}  
We thank Simon Caron-Huot and  Alexander Zhiboedov for discussions. The work of J.R. is supported by Taiwan NSTC grant 112-2811-M-002 -054 -MY2, while Y-t.H. is supported by Taiwan NSTC grant 112-2628-M-002 -003 -MY3. LR is supported by the Beijing Natural Science Foundation International Scientist Project No. IS24014, and the National Natural Science Foundation of China General Program No. 12475070.  A.H. is supported by the
Department of Energy (Grant No. DE-SC0011632) and by the Walter Burke Institute for Theoretical Physics.

\clearpage
\onecolumngrid
\appendix
\pagenumbering{alph}
\section{Adjoint representation matrices}\label{sup: Color}
The projectors for SO(n) adjoint irreps are conveniently represented in terms of single and double traces [35]: 
\footnote{Here the symmetrization and anti-symmetrization is normalized with $\frac{1}{2}$, i.e. $(ab)=\frac{1}{2}(ab+ba)$}
\begin{align}\label{eq: projbasis}
    & \mathbb{P}_1^{i_1 i_2; i_3 i_4} = \frac{2}{n (n - 1)} \text{Tr}[i_1, i_2] \text{Tr}[i_3, i_4]\,,\nonumber\\
    &\mathbb{P}_2^{i_1 i_2; i_3 i_4} = \frac{4}{(n - 2)} \left( \text{Tr}[i_1, i_2, (i_3, i_4)] 
    - \frac{1}{n} \text{Tr}[i_1, i_2] \text{Tr}[i_3, i_4] \right)\,, \nonumber\\
    &\mathbb{P}_3^{i_1 i_2; i_3 i_4} = \frac{2}{3} \left( \text{Tr}[i_1, (i_3] \text{Tr}[i_4), i_2] 
    + \text{Tr}[i_1, i_4, i_2, i_3] \right) 
 - \frac{4}{n - 2} \text{Tr}[i_1, i_2, (i_3, i_4)] 
    + \frac{2}{(n - 1) (n - 2)} \text{Tr}[i_1, i_2] \text{Tr}[i_3, i_4]\,,\nonumber \\
    &\mathbb{P}_4^{i_1 i_2; i_3 i_4} = \frac{1}{3} \left( \text{Tr}[i_1, (i_3] \text{Tr}[i_4), i_2] 
    - 2 \text{Tr}[i_1, i_4, i_2, i_3] \right)\,,\nonumber \\
    &\mathbb{P}_5^{i_1 i_2; i_3 i_4} = \frac{4}{(n - 2)} \text{Tr}[i_1, i_2, [i_3, i_4]]\,, \nonumber\\
    &\mathbb{P}_6^{i_1 i_2; i_3 i_4} =   \text{Tr}[i_1, [i_4]  \text{Tr}[i_3], i_2]
    - \frac{4}{(n - 2)} \text{Tr}[i_1, i_2, [i_3, i_4]]\,.
\end{align}
The matrix $M_{st}$ for adjoint representation given as:

\begin{equation}\label{eq:Mst}
\begin{pmatrix}
\frac{2}{(n-1)n} & \frac{2}{(n-1)n} & \frac{2}{(n-1)n} & \frac{2}{(n-1)n} & \frac{2}{(n-1)n} & \frac{2}{(n-1)n} \\[6pt]
\frac{n+2}{n} & \frac{n^2-8}{2(n-2)n} & \frac{n-4}{(n-2)n} & -\frac{2(n+2)}{(n-2)n} & \frac{(n-4)(n+2)}{2(n-2)n} & -\frac{4}{(n-2)n} \\[6pt]
\frac{(n-3)(n+1)(n+2)}{6(n-1)} & \frac{(n-4)(n-3)(n+1)}{6(n-2)(n-1)} & \frac{11-6n+n^2}{3(n-2)(n-1)} & \frac{(n+1)(n+2)}{3(n-2)(n-1)} & -\frac{(n-3)(n+1)(n+2)}{6(n-2)(n-1)} & -\frac{(n-4)(n+1)}{3(n-2)(n-1)} \\[6pt]
\frac{1}{12}(n-3)(n-2) & \frac{3-n}{6} & \frac{1}{6} & \frac{1}{6} & \frac{n-3}{6} & -\frac{1}{6} \\[6pt]
1 & \frac{n-4}{2(n-2)} & -\frac{1}{n-2} & \frac{2}{n-2} & \frac{1}{2} & 0 \\[6pt]
\frac{1}{4}(n-3)(n+2) & -\frac{n-3}{n-2} & -\frac{n-4}{2(n-2)} & -\frac{n+2}{2(n-2)} & 0 & \frac{1}{2}
\end{pmatrix}
\end{equation}

and similarly for $M_{ut}$:

\begin{equation} \label{eq:Mut}
\begin{pmatrix}
\frac{2}{(n-1)n} & \frac{2}{(n-1)n} & \frac{2}{(n-1)n} & \frac{2}{(n-1)n} & -\frac{2}{(n-1)n} & -\frac{2}{(n-1)n} \\[6pt]
\frac{n+2}{n} & \frac{n^2-8}{2n(n-2)} & \frac{n-4}{(n-2)n} & -\frac{2(n+2)}{(n-2)n} & -\frac{(n-4)(n+2)}{2n(n-2)} & \frac{4}{(n-2)n} \\[6pt]
\frac{(n-3)(n+1)(n+2)}{6(n-1)} & \frac{(n-4)(n-3)(n+1)}{6(n-2)(n-1)} & \frac{n^2-6n+11}{3(n-2)(n-1)} & \frac{(n+1)(n+2)}{3(n-2)(n-1)} & \frac{(n-3)(n+1)(n+2)}{6(n-2)(n-1)} & \frac{(n-4)(n+1)}{3(n-2)(n-1)} \\[6pt]
\frac{(n-3)(n-2)}{12} & \frac{3-n}{6} & \frac{1}{6} & \frac{1}{6} & \frac{3-n}{6} & \frac{1}{6} \\[6pt]
1 & \frac{n-4}{2(n-2)} & -\frac{1}{n-2} & \frac{2}{n-2} & -\frac{1}{2} & 0 \\[6pt]
\frac{(n-3)(n+2)}{4} & -\frac{n-3}{n-2} & -\frac{n-4}{2(n-2)} & -\frac{n+2}{2(n-2)} & 0 & -\frac{1}{2}
\end{pmatrix}
\end{equation}

\section{Vectors $v$ for adjoint} \label{sec:v adj}
The vectors $v$ that satisfies eq.(\ref{eq: NullSpace}) for selected irreps. Firstly, one $\{1,5,6\}$ are possible for three irrep completions. The $v$ that rules out all remaining three irreps are listed in table~\ref{tab:3 reps}. 

\begin{table}[H]
\centering
\small 

\begin{tabular}{|c|c|}
\hline
{Irreps} & $\vec{v}$ \\
\hline
$\{1, 2, 3\}$ & $\left( 1, \frac{2 - n - n^2}{-2 + n}, \frac{2 n + 3 n^2 + n^3}{6 (-2 + n)}, \frac{1}{6} (n - n^2), 0, 0 \right)$ \\
\hline
$\{1, 2, 4\}$ & $\left( 1, \frac{4 - 5 n + n^2}{2 (-2 + n)}, \frac{16 n - 3 n^2 - n^3}{12 (-2 + n)}, \frac{-n + n^2}{12}, 0, 0 \right)$ \\
\hline
$\{1, 2, 5\}$ & $\left( 1, -\frac{2 (-1 + n)}{-2 + n}, \frac{n}{-2 + n}, 0, 0, 0 \right)$ \\
\hline
$\{1, 2, 6\}$ & $\left( 1, \frac{8 - 6 n - 3 n^2 + n^3}{4 (-2 + n)}, \frac{16 n + 10 n^2 - n^3 - n^4}{24 (-2 + n)}, \frac{2 n - 3 n^2 + n^3}{24}, 0, 0 \right)$ \\
\hline
$\{1, 3, 4\}$ & $\left( 1, \frac{4 - 5 n + n^2}{2 (-2 + n)}, \frac{12 n + 5 n^2 - 6 n^3 + n^4}{12 (-2 + n)}, \frac{-3 n + 4 n^2 - n^3}{12}, 0, 0 \right)$ \\
\hline
$\{1, 3, 5\}$ & $\left( 1, \frac{-16 + 16 n + n^2 - n^3}{(-8 + n) (-2 + n)}, \frac{4 n + 3 n^2 - n^3}{6 (-2 + n)}, \frac{2 n - 3 n^2 + n^3}{6 (-8 + n)}, 0, 0 \right)$ \\
\hline
$\{1, 3, 6\}$ & $\left( 1, \frac{-32 + 26 n + 11 n^2 - 5 n^3}{2 (-8 + n) (-2 + n)}, \frac{6 n + 7 n^2 - n^4}{12 (-2 + n)}, \frac{-6 n + 11 n^2 - 6 n^3 + n^4}{12 (-8 + n)}, 0, 0 \right)$ \\
\hline
$\{1, 4, 5\}$ & $\left( 1, \frac{4 - 5 n + n^2}{2 (-2 + n)}, \frac{20 n - 9 n^2 + n^3}{12 (-2 + n)}, \frac{n - n^2}{12}, 0, 0 \right)$ \\
\hline
$\{1, 4, 6\}$ & $\left( 1, \frac{4 - 5 n + n^2}{2 (-2 + n)}, \frac{24 n - 17 n^2 + 6 n^3 - n^4}{12 (-2 + n)}, \frac{3 n - 4 n^2 + n^3}{12}, 0, 0 \right)$ \\
\hline
$\{2, 3, 4\}$ & $\left( 1, -\frac{2 (-2 + n + n^2)}{-4 - n + n^2}, \frac{6 n + 7 n^2 - n^4}{3 (-4 - n + n^2)}, \frac{6 n - 11 n^2 + 6 n^3 - n^4}{6 (-4 - n + n^2)}, 0, 0 \right)$ \\
\hline
$\{2, 3, 5\}$ & $\left( 1, -\frac{4 (-2 + n + n^2)}{-8 + n + n^2}, \frac{8 n + 10 n^2 + n^3 - n^4}{3 (-8 + n + n^2)}, \frac{2 n - 5 n^2 + 4 n^3 - n^4}{6 (-8 + n + n^2)}, 0, 0 \right)$ \\
\hline
$\{2, 3, 6\}$ & $\left( 1, -\frac{2 (8 - 6 n - 3 n^2 + n^3)}{16 - 9 n + n^2}, \frac{2 (-6 n - 7 n^2 + n^4)}{3 (16 - 9 n + n^2)}, \frac{6 n - 11 n^2 + 6 n^3 - n^4}{6 (16 - 9 n + n^2)}, 0, 0 \right)$ \\
\hline
$\{2, 4, 5\}$ & $\left( 1, -\frac{4 (-1 + n)}{-4 + n}, \frac{10 n - n^3}{3 (-4 + n)}, \frac{-2 n + 3 n^2 - n^3}{6 (-4 + n)}, 0, 0 \right)$ \\
\hline
$\{2, 4, 6\}$ & $\left( 1, \frac{(-1 + n) (-8 - 2 n + n^2)}{-8 + n^2}, \frac{48 n + 2 n^2 - 9 n^3 + n^4}{6 (-8 + n^2)}, \frac{-6 n + 11 n^2 - 6 n^3 + n^4}{3 (-8 + n^2)}, 0, 0 \right)$ \\
\hline
$\{2, 5, 6\}$ & $\left( 1, 0, \frac{-4 + n^2}{3}, \frac{2 - 3 n + n^2}{6}, 0, 0 \right)$ \\
\hline
$\{3, 4, 5\}$ & $\left( 1, -\frac{2 (-4 + n) (-1 + n)}{8 - 5 n + n^2}, \frac{-12 n - 5 n^2 + 6 n^3 - n^4}{3 (8 - 5 n + n^2)}, -\frac{(-3 n + n^2) (2 - 3 n + n^2)}{6 (8 - 5 n + n^2)}, 0, 0 \right)$ \\
\hline
$\{3, 4, 6\}$ & $\left( 1, \frac{-2 + n + n^2}{2}, \frac{-6 n - 7 n^2 + n^4}{12}, \frac{6 n - 11 n^2 + 6 n^3 - n^4}{12}, 0, 0 \right)$ \\
\hline
$\{3, 5, 6\}$ & $\left( 1, -\frac{2 (-4 + n^2)}{-8 + n}, 0, \frac{6 n - 5 n^2 + n^3}{2 (-8 + n)}, 0, 0 \right)$ \\
\hline
$\{4, 5, 6\}$ & $\left( 1, -1 + n, \frac{-3 n + n^2}{2}, 0, 0, 0 \right)$ \\
\hline
\end{tabular}
\caption{$v$ that force $G=0$ when UV only has three irreps. Adjoint representation case.}
\label{tab:3 reps}
\end{table}

It is possible for $v$s to exist for particular subsets with specific $n$. Note that even though SO($n$)$\subset$SO($n'$) for $n<n'$, the subset of irreps for SO($n'$) generally span a larger set in SO($n$). An example is for $n=4$, we find that $\{1, 2, 4, 6\}$ and $\{1, 3, 4, 5\}$ are ruled out, with the vectors given in table~\ref{tab:4 reps}.

\begin{table}[h!]
\centering
\begin{tabular}{|c|c|}
\hline
Irreps & $\vec{v}$   \\
\hline
$\{1, 2, 4, 6\}$ & $( 1 , 0 , -2 , 1 , 0 , 0 )$   \\
\hline
$\{1, 3, 4, 5\}$ & $( 1 , 0 , 0 , -1 , 0 , 0 )$    \\
\hline
\end{tabular}
\caption{$v$ that force $G=0$ when UV only has four irreps. Adjoint representation case, $n=4$.}
\label{tab:4 reps}
\end{table}

The vectors that rules out single irreps completions for adjoint matter coupled to gauge field. Only adjoint irreps are possible for single irrep completions. 

\begin{table}[h!]
\centering
\begin{tabular}{|c|c|}
\hline
Representations & $\vec{v}$ \\
\hline
\{1\} & $(0, 0, 0, 0, 1, -1)$ \\
\hline
\{2\} & $\left(0, 0, 0, 0, 1, \frac{-8 - 2 n + n^2}{8}\right)$ \\
\hline
\{3\} & $\left(0, 0, 0, 0, 1, \frac{6 + n - n^2}{2 (-4 + n)}\right)$ \\
\hline
\{4\} & $(0, 0, 0, 0, 1, -3 + n)$ \\

\hline
\{6\} & $(0, 0, 0, 0, 1, 0)$ \\
\hline
\end{tabular}
\caption{$v$ that leads to an in consistency with a $t$-channel massless pole in the adjoint representation. }
\label{tab:1subtr adj}
\end{table}

\section{Improved dispersion relation} \label{sec:impr disp}

A clever way to ``have ones cake and eat it'' was introduced in~\cite{Caron-Huot:2021rmr} where one considers a judicious sum of eq.(\ref{eq:disp3}) and its derivative in $s$ such that only a finite number of EFT coefficients are left, for which non-trivial bounds can be derived. We will proceed in similar fashion, with special attention to the non-trivial color structures. Schematically the improved dispersion relation is written as
\begin{equation}
\label{eq:disp improved sch}
\boldsymbol{\mathcal{C}}^{\rm improved}_{\text{low E},t} = \left<\boldsymbol{\mathcal{C}}^{\rm improved}_{\text{high E},t}[m^2,J,R]\right> \,,
\end{equation}

where on the LHS of the above purely consists of EFT data (finite number of couplings), while the RHS can be expressed in terms of dispersive integral using eq.(\ref{eq:disp5}). We wrote this in vector form, where each component is a different low energy color structure basis element, with the basis for adjoint representation given in \eqref{cDef}. We call the high energy side of eq.(\ref{eq:disp5}) $\bar{A}^{i_1i_2i_3i_4}$, which can be expressed as:
\be\label{eq:Atil}
\bar{A}^{i_1i_2i_3i_4}(s,t)=\sum_R\left< \frac{su \mathbb{G}_{J}^{(D)}(1+\frac{2t}{m^2})}{m^2 + t} \left(\frac{\mathbb{P}_R^{i_1i_2;i_3i_4}}{(m^2-s)}+ \frac{\mathbb{P}_R^{i_1i_3;i_2i_4}}{(m^2+t+s)}\right)\right>_R.
\ee
 Here 
\be\label{sum}
\left<\cdots\right>_R=\frac{1}{\pi}\sum_{J} \, n_{J}^{(D)}\int^\infty_{M^2} \frac{dm^2}{m^2}\; m^{D{-}4} \rho_{J,R}(m^2)\cdots\,,
\ee
where $\rho_{J,R}(m^2)=\text{Im} [f_{J, R}(m^2)]$ and one now has to sum over the irreps $R$, $J$ must be even for symmetric $R$, and odd for antisymmetric.

Define $\tilde{A}^{i_1 i_2 i_3 i_4}(s,t)\equiv su \bar{A}^{i_1 i_2 i_3 i_4}(s,t)$, which is finite in the forward limit. Using the notation 
$$\tilde{A}^{(n,m)i_1 i_2 i_3 i_4}(s^*,t^*)\equiv\partial_s^n \partial_t^m \tilde{A}^{(n,m)i_1 i_2 i_3 i_4}(s,t)\bigg|_{s=s^*, t=t^*}\,,$$
we have:
\be\label{eq: EFTId} 
\begin{split}
&\tilde{A}^{(1,0) i_1 i_2 i_3 i_4}(0,t)={-}B_{\text{poly}}^{(1,0)i_1 i_2 i_3 i_4}(0,t){+}\frac{1}{t}B_{\text{poly}}^{(1,0)i_1 i_2 i_3 i_4}(0,t){-}\frac{1}{t}B_{\text{poly}}^{(1,0)i_1 i_2 i_3 i_4}(-t,t){+}8\pi G  \mathbb{P}_1^{i_2i_3;i_4i_1},
\end{split}
\ee
where the polynomial part of the amplitude is
\begin{eqnarray}
    \label{eq: PolyDef}
  \text{fundamental}:\quad   B_{\text{poly}}^{i_1 i_2 i_3 i_4}(s,t)&=&B(s, t)\delta^{i_1 i_2}\delta^{i_3 i_4}+ B(u, s)\delta^{i_1 i_3}\delta^{i_2 i_4}+B(t, u)\delta^{i_1 i_4}\delta^{i_2 i_3}\nonumber\\
  \text{adjoint}:\quad   B_{\text{poly}}^{i_1 i_2 i_3 i_4}(s,t)&=&\sum_{\sigma \in S_3} \text{Tr}[i_1 \sigma(i_2)\sigma(i_3)\sigma(i_4)]  B_1(1,\sigma(2),\sigma(3),\sigma(4))\nonumber\\&&+ B_2(s,t) \text{Tr}[i_1i_2]\text{Tr}[i_3i_4]+B_2(t,u) \text{Tr}[i_1 i_4]\text{Tr}[i_2 i_3]+B_2(u,t) \text{Tr}[i_1i_3]\text{Tr}[i_2 i_4]\,.\nonumber\\
\end{eqnarray}
In our convention $\text{Tr}(T^{i_1}T^{i_2})=\delta^{i_1i_2}$. The "$B$" functions are polynomials in Mandelstam variables that respect the symmetry of their associated color factors:
\begin{eqnarray}
    \label{eq:BF}
    B(s,t) &=& \sum_{k,q \leq k, q \in \text{even} }G_{k q}\; s^{k-{q}}(t-u)^{q}\,\nonumber\\
    B_1(1234)&=&\sum_{k,q \leq k, q \in \text{even} } g_{k q}\; t^{k-q} (s-u)^q,\nonumber\\
B_2(s,t)&=&\sum_{k,q \leq k, q \in \text{even} } G_{k q}\; s^{k-q} (t-u)^q\,.
\end{eqnarray}
We will use $G_{kq}$ to denote the double trace Wilson coefficient and $g_{kq}$ to denote the single trace coefficients. 

Using crossing symmetry we can write $B^{i_1 i_2 i_3 i_4}(-t,t)=B^{i_1 i_3 i_2 i_4}(0,t)$. Then we expand the equation above in Taylor series:

\begin{align}\label{eq:A01}
\tilde{A}^{(1,0)i_1 i_2 i_3 i_4}(0,t)=8\pi G  \mathbb{P}_0^{i_2i_3;i_4i_1}+\sum_{n=0}\left(-\frac{B_{\text{poly}}^{(1,n)i_1 i_2 i_3 i_4}(0,0)}{n!}+\frac{B_{\text{poly}}^{(0,n+1)i_1 i_2 i_3 i_4}(0,0)}{(n+1)!}-\frac{B_{\text{poly}}^{(0,n+1)i_1 i_3 i_2 i_4}(0,0)}{(n+1)!}\right)t^n\,.
\end{align}

Let us consider the first two non-zero terms in this expansion
\be\label{eq: B2resum} 
\begin{split}
&\sum_{n=0}^2 \left(-\frac{B_{\text{poly}}^{(1,n)i_1 i_2 i_3 i_4}(0,0)}{n!}+\frac{B_{\text{poly}}^{(0,n+1)i_1 i_2 i_3 i_4}(0,0)}{(n+1)!}-\frac{B_{\text{poly}}^{(0,n+1)i_1 i_3 i_2 i_4}(0,0)}{(n+1)!}\right)t^n\equiv t c^{i_1 i_2 i_3 i_4}(t),
\end{split}
\ee
For adjoint scalars $c^{i_1 i_2 i_3 i_4}(t)$ is given as:
\be\label{cDef} 
c^{i_1 i_2 i_3 i_4}(t)=
 \begin{pmatrix} 2 t (g_{30}-g_{32})-g_{20}-g_{22}\\t (g_{30}-3 g_{32})-g_{20}-g_{22}\\-4 (t
   g_{32}+g_{22})\\t (G_{30}-3 G_{32})-G_{20}-G_{22}\\-4 (t G_{32}+G_{22})\\2 t (G_{30}-G_{32})-G_{20}-G_{22} \end{pmatrix} \cdot  \begin{pmatrix} \text{Tr}[i_1,i_2,i_3,i_4]\\ \text{Tr}[i_1,i_4,i_2,i_3]\\  \text{Tr}[i_1,i_3,i_4,i_2] \\  \text{Tr}[i_1i_4] \text{Tr}[i_2i_3]\\  \text{Tr}[i_1i_3]\text{Tr}[i_2i_4]\\  \text{Tr}[i_1i_3]\text{Tr}[i_2i_4] \end{pmatrix}\;.
\ee 
For fundamental scalars we only have the last three rows. One sees that one has $4$ single trace and $4$ double trace coefficients remaining. One can further consider linear combinations such that only two are left for each.
For $n>2$, using crossing we can rewrite it as:
\be\label{eq: B3resumm} 
\begin{split}
&\sum_{n=3}^\infty \left(-\frac{B_{\text{poly}}^{(1,n)i_1 i_2 i_3 i_4}(0,0)}{n!}+\frac{B_{\text{poly}}^{(0,n+1)i_1 i_2 i_3 i_4}(0,0)}{(n+1)!}-\frac{B_{\text{poly}}^{(0,n+1)i_1 i_3 i_2 i_4}(0,0)}{(n+1)!}\right)t^n=\\
=&\ \tilde{A}^{(0,1)i_1 i_4 i_3 i_2}(t,0)-\tilde{A}^{(0,1)i_1 i_4 i_3 i_2}(t,0)|_{\mathcal{O}(2)}+\\
&+\frac{1}{t}\bigg(-\tilde{A}^{i_1 i_4 i_3 i_2}(t,0)+\tilde{A}^{i_1 i_4 i_3 i_2}(t,0)|_{\mathcal{O}(3)}+\tilde{A}^{i_4 i_2 i_1 i_3}(t,0)-\tilde{A}^{i_4 i_2 i_1 i_3}(t,0)|_{\mathcal{O}(3)} \bigg)\,.
\end{split}
\ee
where we use the shorthand notation $\tilde{A}(t,0)|_{\mathcal{O}(n)}$ to represent expansion in $t$ to the $n$-th order. Combining eq.(\ref{eq: B2resum}) and eq.(\ref{eq: B3resumm}) into  \eqref{eq: EFTId} we have the following identity:
\be \label{eq:disp improved}
\boxed{\begin{split}
&c^{i_1i_2i_3i_4}(t)+8\pi G  \frac{\mathbb{P}_0^{i_2i_3;i_4i_1}}{t}=\\
&\frac{1}{t}\Bigg[\tilde{A}^{(1,0)i_1i_2i_3i_4}(0,t)-
\tilde{A}^{(0,1)i_1 i_4 i_3 i_2}(t,0)+\tilde{A}^{(0,1)i_1 i_4 i_3 i_2}(t,0)|_{\mathcal{O}(2)}-\\&\frac{1}{t}\bigg(-\tilde{A}^{i_1 i_4 i_3 i_2}(t,0)+\tilde{A}^{i_1 i_4 i_3 i_2}(t,0)|_{\mathcal{O}(3)}+\tilde{A}^{i_4 i_2 i_1 i_3}(t,0)-\tilde{A}^{i_4 i_2 i_1 i_3}(t,0)|_{\mathcal{O}(3)} \bigg)\Bigg]
\end{split}}\,.
\ee

\section{Choices of parameters and numerical results}
Our numerical optimization procedure follows \cite{Caron-Huot:2021rmr}. We smeared \eqref{eq:disp improved} by integrating it against $(1-p)p^n$, where $t=-p^2$.
Overall we have three regions in the spectrum $(m,J)$ space in which we need to impose the positivity of $\rho_{J,R}(m^2)$ in \eqref{sum}. Here we give details of our choices of parameters in each of them
\begin{enumerate}
  \item Finite $m>M$ and $J<J_{max}$.  Here we imposed positivity of  at discretized 100 values of $x=1-\frac{M}{\mu} $ between 0 and 1, later plotting the functionals and adding more points at the regions in which they are negative. The maximum spin was $J_{max}=40$ at which seemed to be sufficient for the convergence.
  \item $m\rightarrow\infty$, $J\rightarrow\infty$ and finite $b=\frac{2J}{m}<b_{max}$.  Here we imposed positivity at discretized 400 values of $b$ between $0$ and $b_{max}=40$, later plotting the functionals and adding more points at the regions in which they are negative.
  \item Large $b>b_{max}$. In this region we do not need to discretize, the positivity statement becomes a requirement that a certain matrix has to be positive semi-definite (see \cite{Caron-Huot:2021rmr}).
\end{enumerate}

In every of these regions we analytically calculated $p$ integrals up to $p^9$. To take into account sufficient number of null constraints and also probe the higher couplings that are not in the improved dispersion relations we added the forward limit of the higher order $s$ and $t$ terms of \eqref{eq:disp5}, where the graviton pole does not contribute. This way we could probe EFT couplings up to $G_{10 q}$ and $g_{10 q}$ and also all null constraints that are obtained by combining the forward limit dispersion relations.

\twocolumngrid
\bibliography{refs}

\end{document}